\documentclass[12pt,a4paper]{article}
\usepackage{amsmath}
\usepackage{amssymb}
\usepackage{color}
\usepackage{graphicx}
\usepackage{float}
\usepackage{cite}
\usepackage{dcolumn} 
\newcolumntype{.}{D{.}{.}{1}}
\usepackage[ruled, vlined, linesnumbered]{algorithm2e}
\usepackage{setspace}

\usepackage{algpseudocode}              

\usepackage{mathrsfs}
\usepackage{booktabs,threeparttable}
\usepackage{mhchem}
\usepackage{braket}
\usepackage{dsfont}
\usepackage{mathtools}
\usepackage{paralist}
\usepackage{comment}
\usepackage{xcolor}
\usepackage[normalem]{ulem}

\begin{document}
\begin{center}

\vspace*{0.5cm}

{\Large\bf
Developing Orbital-Dependent Corrections for the
Non-Additive Kinetic Energy in Subsystem Density
Functional Theory
}

\vspace{1.5cm}

{\large Larissa Sophie Eitelhuber and Denis G. Artiukhin\footnote{Email: denis.artiukhin@fu-berlin.de} \\[2ex]
}

\vspace{1.5cm}
Institut f\"ur Chemie und Biochemie, Freie Universit\"at Berlin, \\ Arnimallee 22, 14195 Berlin, Germany \\[1ex]
\vspace{0.5cm}

\end{center}

\vfill

\begin{tabbing}
Date:   \quad\= \today \\
\end{tabbing}

\newpage

\begin{abstract}

We present a novel route to constructing cost-efficient semi-empirical approximations for the non-additive kinetic energy in subsystem density functional theory. The developed methodology is based on the use of Slater determinants composed of non-orthogonal Kohn--Sham-like orbitals for the evaluation of kinetic energy expectation values and the expansion of the inverse molecular-orbital overlap matrix into a Neumann series. Applying these techniques, we derived and implemented a series of orbital-dependent approximations for the non-additive kinetic energy, which are employed  self-consistently. Our proof-of-principle computations demonstrated quantitatively correct results for potential energy curves and electron densities and hinted on the applicability of the introduced empirical parameters to different types of molecular systems and intermolecular interactions. We therefore conclude that the presented study is an important step towards constructing accurate and efficient orbital-dependent approximations for the non-additive kinetic energy applicable to large molecular systems.

\end{abstract}

\newpage
\clearpage

\section{Introduction} \label{sec:intro}

Subsystem Density Functional Theory (sDFT)~\cite{senatore1986,johnson1987,cortona1991} is based on the commonly used Kohn--Sham Density Functional Theory (KS-DFT) and adopts the idea of partitioning the total molecular system into subsystems based on the electron density. This approach can provide a very favorable computational scaling allowing one to compute large molecular systems composed of up to a few thousand atoms~\cite{konig2013}. 
However, due to the non-additive nature of the exchange--correlation (XC) and kinetic energies, the density partitioning gives rise to 
new terms in the sDFT energy expression. 
As long as the XC energy is given by a pure functional of the density, the corresponding non-additive XC contribution is trivial to compute.
Unfortunately, the orbital-dependent non-additive kinetic energy expression in a monomer basis is unknown and requires an additional approximation to be made~\cite{jacob2014}.

Several different strategies were developed over the last decades to account for this energy contribution or, alternatively, to avoid the problem altogether. Among those are decomposable approximations based on the use of explicit density-dependent kinetic energy functionals (for examples, see Ref.~\cite{jacob2014}), the projection-based embedding theory~\cite{huzinaga1971,manby2012,khait2012} enforcing external orthogonality between subsystem orbitals and ensuring that the non-additive kinetic energy is equal to zero, and the potential reconstruction technique~\cite{Roncero2008,fux2010,manby_miller2010,huang2011}, which is employed to obtain accurate embedding potentials. Unfortunately, the use of explicit kinetic energy functionals is associated with limitations such as the inability to describe strongly interacting molecules and to cut through covalent bonds, whereas projection-based embedding and potential reconstruction techniques often lead to a large increase in the computational cost of sDFT. Therefore, the problem of accurately approximating non-additive kinetic energy contributions in a cost-efficient way persists.
For more information on this topic, we refer to the recent review of sDFT in Ref.~\cite{jacob2024}.

As opposed to projection-based embedding, which enforces external orthogonality between subsystems, an approximate strategy employing Slater determinants composed of non-orthogonal Kohn--Sham-like molecular orbitals (MOs) for the direct evaluation of expectation values of quantum operators was demonstrated within 
the Frozen-Density Embedding Diabatization (FDE-diab) technique~\cite{pavanello2011,pavanello2013}.
This approach was originally developed by Pavanello et al.\ and received an increasing attention over the last years (see Refs.~\cite{arti2018,arti2020a,arti2021,eschenbach2021,eschenbach2022,eschenbach2022b}).
Although not being formally exact, it was successfully used for electron- and hole-transfer simulations~\cite{solo2014,ramos2015}, computations of spin-densities~\cite{arti2018,arti2020a,arti2021}, and isotropic components of hyperfine coupling constants~\cite{eschenbach2022}. To the best of our knowledge and somewhat surprisingly, this strategy has never been tested in the context of sDFT for computations of kinetic energy contributions. Being inspired by the previous success of FDE-diab, we make the first important step towards filling this gap by developing orbital-dependent approximations for the non-additive kinetic energy, which are based on non-orthogonal Kohn--Sham-like MOs.
In this regard, our main priority does not lie in the construction of a formally exact theory, as opposed to many existing approaches, but rather in creating an alternative route to inexpensive, practical, and 
fully self-consistent sDFT computations applicable to large molecular systems.

This work is organized as follows. The formal theory behind the sDFT method as well as derivations of orbital-dependent approximations for the non-additive kinetic energy are outlined in Sec.~\ref{sec:theory}. Computational details for numerical tests conducted are given in Sec.~\ref{sec:comput_details}.
Subsequently, assessments of approximations made and proof-of-principle computations employing the new approximations are presented in Sec.~\ref{sec:results}. Conclusions to the results and associated discussions are given in Sec.~\ref{sec:conclusions}.

\section{Theory} \label{sec:theory}

In the following, we briefly outline the theory behind sDFT in Sec.~\ref{sec:sDFT}. For more details on this topic, the interested reader is referred to reviews in Refs.~\cite{jacob2014,jacob2024}. Subsequently, new approximations for the non-additive kinetic energy are described in Sec.~\ref{sec:nadd_kin}.

\subsection{Subsystem Density Functional Theory} \label{sec:sDFT}

As mentioned above, the central idea of sDFT~\cite{senatore1986,johnson1987,cortona1991} lies in the partitioning of the electron density $\rho (\vec{r}\,)$ for the total molecular system into subsystem densities. For the sake of simplicity, we only consider here the case of the total system being composed of two subsystems $A$ and $B$. Note that the generalization of our new approach to multiple subsystems is not completely trivial and could require additional approximations to be made.
The corresponding density partitioning for two subsystems reads
\begin{equation} \label{eq:dens_partitioning}
	\rho (\vec{r}\,) = \rho_A (\vec{r}\,) + \rho_B (\vec{r}\,).
\end{equation}
Each subsystem density $\rho_I (\vec{r}\,)$, where $I = A$ or $B$, is computed from corresponding $N_I$ orthonormal occupied Kohn--Sham-like MOs $\{\psi_{i}^I \}_{i=1}^{N_I}$,
\begin{equation} \label{eq:electron_density}
	\rho_I (\vec{r}\,) = \sum_{i = 1}^{N_I}  |\psi_i^I (\vec{r}\,) |^2,
\end{equation}
which describe a reference non-interacting subsystem $I$ of electrons. 
If the sets of MOs $\{\psi_{i}^A \}_{i=1}^{N_A}$ and $\{\psi_{i}^B \}_{i=1}^{N_B}$ are mutually orthonormal, the non-interactive kinetic energy of the total molecular system $T_s [\{ \psi_i \}]$ is equal to the sum of subsystem contributions $T_s [\{ \psi_i^A \}]$ and $T_s [\{ \psi_i^B \}]$, which are computed from the corresponding sets of orbitals $\{\psi_{i}^A \}_{i=1}^{N_A}$ and $\{\psi_{i}^B \}_{i=1}^{N_B}$, respectively.
In this case, the set of MOs for the total molecular system $\{ \psi_i \}_{i=1}^{N_A + N_B}$ is simply a union of the subsystem MO sets, i.e., $\{ \psi_i \}_{i=1}^{N_A + N_B} = \{\psi_{i}^A \}_{i=1}^{N_A} \cup \{\psi_{i}^B \}_{i=1}^{N_B}$. However, in practical computations mutual orthonormality is often not enforced and the orthonormal set $\{ \psi_i \}_{i=1}^{N_A + N_B}$ is unknown. As a result, the kinetic energy of the total molecular system $T_s [\{ \psi_i \}]$ is not available, but is formally given as
\begin{equation} \label{eq:non_additive_kin_energy}
	T_s [\{ \psi_i  \}] = T_s [\{ \psi_i^A \}] + T_s [\{ \psi_i^B \}] + T_s^{\mathrm{nad}} [\rho_A, \rho_B ],
\end{equation}
where $T_s^{\mathrm{nad}} [\rho_A, \rho_B ]$ is the non-additive kinetic energy correction. The term $T_s^{\mathrm{nad}} [\rho_A, \rho_B ]$ cannot be evaluated directly from Eq.~(\ref{eq:non_additive_kin_energy}) and, therefore, is often approximated with explicit functionals of density.

Analogously, a non-additive correction term appears in the expression for the XC energy of the total molecular system $E_{\mathrm{XC}} [\rho]$, 
\begin{equation} \label{eq:non_additive_XC_energy}
	E_{\mathrm{XC}} [\rho] = E_{\mathrm{XC}} [\rho_A] + E_{\mathrm{XC}} [\rho_B] + E_{\mathrm{XC}}^{\mathrm{nad}} [\rho_A, \rho_B ].
\end{equation}
Here, $E_{\mathrm{XC}} [\rho_A]$, $E_{\mathrm{XC}} [\rho_B]$, and $E_{\mathrm{XC}}^{\mathrm{nad}} [\rho_A, \rho_B]$ are 
XC energy contributions from subsystems $A$ and $B$ as well as the non-additive XC correction, respectively. As long as all terms from Eq.~(\ref{eq:non_additive_XC_energy}) are given as pure functionals of density, the total XC energy $E_{\mathrm{XC}} [\rho]$ as well as subsystem contributions $E_{\mathrm{XC}} [\rho_A]$ and $E_{\mathrm{XC}} [\rho_B]$
can be computed exactly. This is made possible by the fact that the density $\rho (\vec{r}\,)$ of the total molecular system is equal to the sum of subsystem densities $\rho_A (\vec{r}\,)$ and $\rho_B (\vec{r}\,)$, as shown in Eq.~(\ref{eq:dens_partitioning}), and can easily be computed. Therefore, the non-additive XC energy $E_{\mathrm{XC}}^{\mathrm{nad}} [\rho_A, \rho_B]$ could be evaluated from Eq.~(\ref{eq:non_additive_XC_energy}) without the need to introduce any new approximations.

With both non-additive energy terms being defined, the energy for the total molecular system can be represented as
\begin{multline} \label{eq:sdft_energy_expression}
	E [\rho] = T_s [\{ \psi_i^A \}] + T_s [\{ \psi_i^B \}] + U_{\mathrm{ext}} [\rho] +   J [\rho] + E_{\mathrm{XC}} [\rho_A] + E_{\mathrm{XC}} [\rho_B] \\   
	+ T_s^{\mathrm{nad}} [\rho_A, \rho_B]
	+ E^{\mathrm{nad}}_{\mathrm{xc}}  [\rho_A, \rho_B] +  U_{\mathrm{nuc}},
\end{multline}
where $U_{\mathrm{ext}} [\rho]$ is the external potential, 
$J [\rho]$ is the classical Coulomb electronic repulsion, 
and $U_{\mathrm{nuc}}$ is the nucleus--nucleus repulsion, all being additive quantities and known from standard KS-DFT. To find the ground state energy, the expression from Eq.~(\ref{eq:sdft_energy_expression}) has to be minimized with respect to both subsystem densities $\rho_A (\vec{r}\,)$ and $\rho_B (\vec{r}\,)$. In practice, this can be achieved by performing a series of constrained minimizations. First, the energy $E [\rho]$ is minimized with respect to an electron density of a single subsystem (referred to as ``active'') while keeping another subsystem (called ``environment'') density fixed. Subsequently, the roles of active and environment subsystems are exchanged and the minimization procedure is repeated until the full relaxation of the total electron density. This constrained minimization approach is known as Frozen Density Embedding (FDE)~\cite{wesolowski1993}, whereas the iterative minimization procedure is called freeze-and-thaw cycles~\cite{weso1996}.

If we regard subsystem $A$ as active and minimize $E [\rho]$ with respect to $\rho_A (\vec{r}\,)$, we obtain the Kohn--Sham Equations with Constrained Electronic Density (KSCED)~\cite{wesolowski1993,weso2006},
\begin{equation} \label{eq:ksced}
\left[  \hat{t} + \upsilon^{(A)}_{\mathrm{eff}} [\rho_A] (\vec{r}\,) 
+ \upsilon^{(A)}_{\mathrm{emb}} [\rho_{A},\rho_B] (\vec{r}\,)
\right] \psi_i^A (\vec{r}\,)  = \varepsilon_i^A  \psi_i^A (\vec{r}\,).
\end{equation}
Here, $\hat{t}$ denotes the one-electron kinetic energy operator $- \nabla^2/2  $. The term $\upsilon^{(A)}_{\mathrm{eff}} [\rho_A] (\vec{r}\,)$ is the effective potential, 
\begin{equation} \label{eq:eff_pot}
	\upsilon^{(A)}_{\mathrm{eff}} [\rho_A] (\vec{r}\,) = \upsilon^{(A)}_{\mathrm{nuc}} (\vec{r}\,) +  \upsilon_{\mathrm{Coul}} [\rho_A] (\vec{r}\,) + \upsilon_{\mathrm{xc}} [\rho_A] (\vec{r}\,),
\end{equation}
which is similar to that from standard KS-DFT but is defined for active subsystem $A$. It includes the nuclear $\upsilon^{(A)}_{\mathrm{nuc}} (\vec{r}\,)$, Coulomb $\upsilon_{\mathrm{Coul}} [\rho_A] (\vec{r}\,)$, and XC 
$\upsilon_{\mathrm{xc}} [\rho_A] (\vec{r}\,)$
potential contributions. The new term $\upsilon^{(A)}_{\mathrm{emb}} [\rho_A, \rho_B] (\vec{r}\,)$, which is not present in KS-DFT, is the embedding potential accounting for the interaction between subsystems. It is given as
\begin{equation} \label{eq:embedding_potential_fde_theory}
	\upsilon^{(A)}_{\mathrm{emb}} [\rho_A,\rho_B] (\vec{r}\,)  =
	\upsilon^{(B)}_{\mathrm{nuc}} (\vec{r}\,) +
	\upsilon_{\mathrm{Coul}} [\rho_B ] (\vec{r}\,)
	+ \upsilon^{\mathrm{nad}}_{\mathrm{xc}} [\rho_A,\rho_B] (\vec{r}\,) + 
	\upsilon^{\mathrm{nad}}_{\mathrm{kin}} [\rho_A,\rho_B] (\vec{r}\,).
\end{equation}
Here, $\upsilon^{\mathrm{nad}}_{\mathrm{kin}} [\rho_A,\rho_B] (\vec{r}\,)$ is the functional derivative of the non-additive kinetic energy $T_s^{\mathrm{nad}} [\rho_A, \rho_B]$ with respect to  the active subsystem density $\rho_A (\vec{r}\,)$,
\begin{equation} \label{eq:kin_func_der}
\upsilon^{\mathrm{nad}}_{\mathrm{kin}} [\rho_A,\rho_B] (\vec{r}\,) = \frac{\delta T_s^{\mathrm{nad}} [\rho_A, \rho_B]}{\delta \rho_A  (\vec{r}\,)} = 
\frac{\delta T_s [\rho ]}{\delta \rho (\vec{r}\,)}  
- 	\frac{\delta T_s [\rho_A ]}{\delta \rho_A (\vec{r}\,)}.
\end{equation}
The non-additive XC potential $\upsilon^{\mathrm{nad}}_{\mathrm{xc}} [\rho_A,\rho_B] (\vec{r}\,)$ is defined analogously as the functional derivative of the non-additive XC energy $E^{\mathrm{nad}}_{\mathrm{xc}}  [\rho_A, \rho_B]$.

\subsection{Non-Additive Kinetic Energy Correction} \label{sec:nadd_kin}

In order to derive new orbital-dependent expressions for the non-additive kinetic energy, we start from approximating the total system non-interactive kinetic energy $T_s$ as
\begin{equation} \label{eq:main_assumption}
	T_s \approx \frac{\braket{\Phi | \hat{T} | \Phi   }}{\braket{\Phi | \Phi}},
\end{equation}
where $\Phi$ is a Slater determinant composed of two sets of Kohn--Sham-like MOs $\{\psi_{i}^A \}_{i=1}^{N_A}$ and $\{\psi_{i}^B \}_{i=1}^{N_B}$, which are not mutually orthogonal, and $\hat{T}$ is the operator of electronic kinetic energy of the total molecular system. This expression could further be re-written in terms of MOs as~\cite{lowdin1955,mayer2003}
\begin{equation} \label{eq:kin_en_nonorth}
	T_s \approx \sum_{i,j=1}^{N_A + N_B} \braket{ \phi_i | \hat{t} | \phi_j } (\mathbf{S}^{-1})_{ji}.
\end{equation}
Here, $\{\phi_i \}_{i=1}^{N_A + N_B} = \{\psi_{i}^A \}_{i=1}^{N_A} \cup \{\psi_{i}^B \}_{i=1}^{N_B}$ is a non-orthogonal set of MOs and
$\mathbf{S}^{-1}$ is the inverse of the MO overlap matrix $\mathbf{S}$ containing elements $S_{ij} = \braket{ \phi_i | \phi_j }$. Note that both the original MO overlap matrix $\mathbf{S}$ and its inverse $\mathbf{S}^{-1}$ are real symmetric matrices, which means that $(\mathbf{S}^{-1})_{ji} = (\mathbf{S}^{-1})_{ij}$. Same holds for kinetic energy integrals, i.e., $\braket{ \phi_i | \hat{t} | \phi_j } = \braket{ \phi_j | \hat{t} | \phi_i }$. However, expressions presented in this work do not account for these properties and are derived in a more general case of complex-valued non-symmetric matrices. This is merely a matter of convenience when deriving functional derivatives as is done later in the text.

The calculation of the inverse overlap matrix $\mathbf{S}^{-1}$ scales as 
$\mathcal{O} ([N_A+N_B]^3)$ with respect to the number of MOs and, therefore, is rather expensive. However, less expensive approximate expressions for $T_s$ from Eq.~(\ref{eq:kin_en_nonorth}) can be obtained assuming that 
$\mathbf{S}^{-1}$ can be expanded into the Neumann series~\cite{neumann1877,renardy2004},
\begin{equation} \label{eq:overlap_neumann}
	\mathbf{S}^{-1} = \sum_{n=0}^{\infty} (\mathbf{I} - \mathbf{S})^{n},
\end{equation}
where $\mathbf{I}$ is the identity matrix. Convergence of this series is further discussed and analyzed in Sec.~\ref{sec:results_neumann}. Note that, when being applied to the inverse MO overlap matrix $\mathbf{S}^{-1}$, the expression from Eq.~(\ref{eq:overlap_neumann}) is also known as the L\"owdin expansion~\cite{loewdin1950,loewdin1956}.

Substituting Eq.~(\ref{eq:overlap_neumann}) into Eq.~(\ref{eq:kin_en_nonorth}), we obtain an expression for the kinetic energy $T_s$, which takes a form of the series, 
\begin{equation} \label{eq:kin_energy_expanded}
	T_s \approx T_s^{(0)} + T_s^{(1)} + T_s^{(2)} + \dots = \sum_{n=0}^{\infty} T_s^{(n)}.
\end{equation}	
It can be shown that the first three terms of this new expansion are equal to
\begin{equation} \label{eq:T0_energy}
	T_s^{(0)} =  \sum_{I=A,B} \sum_{i=1}^{N_I} \braket{ \psi_i^I | \hat{t} | \psi_i^I } = T_s [\{ \psi_i^A \}] + T_s [\{ \psi_i^B \}],
\end{equation}
\begin{equation} \label{eq:T1_energy}
	T_s^{(1)} = - \sum_{i=1}^{N_A} \braket{ \psi_i^A | \hat{t} \hat{\rho}_B + \hat{\rho}_B \hat{t} | \psi_i^A }  ,
\end{equation}	
\begin{equation} \label{eq:T2_energy}
	T_s^{(2)} = \frac{1}{2} \sum_{i=1}^{N_A} \braket{ \psi_i^A |  \hat{t} \hat{\rho}_A \hat{\rho}_B +  \hat{\rho}_B \hat{\rho}_A \hat{t} + 2 \hat{\rho}_B \hat{t} \hat{\rho}_B | \psi_i^A }.
\end{equation}
Here, $\hat{\rho}_I$ are projection operators given by
\begin{equation}
	\hat{\rho}_I = \sum_{i=1}^{N_I} \ket{\psi_i^I} \bra{\psi_i^I}.
\end{equation}
More detailed derivations of these expressions can be found in Sec.~S1 of the Supporting Information (SI).

As one can see, the zero-order expansion term $T_s^{(0)}$ from Eq.~(\ref{eq:T0_energy}) is equal to the sum of subsystem kinetic energies $T_s [\{ \psi_i^A \}]$ and $T_s [\{ \psi_i^B \}]$, which are equivalent to those from the sDFT energy expression in Eq~(\ref{eq:sdft_energy_expression}). Therefore, using the definition of the non-additive kinetic energy $T_s^{\mathrm{nad}}$ from Eq.~(\ref{eq:non_additive_kin_energy}), we can approximate $T_s^{\mathrm{nad}}$ as
\begin{equation} \label{eq:final_non_additive_correction}
	T_s^{\mathrm{nad}} \approx T_s - T_s^{(0)} = T_s^{(1)} + T_s^{(2)} + \dots = \sum_{n=1}^{\infty} T_s^{(n)}.
\end{equation}
This expression is the central assumption analyzed in this work as it provides a route to developing orbital-dependent approximations for $T_s^{\mathrm{nad}}$ as opposed to commonly employed density-based kinetic energy functionals.

It is also interesting to note that the first-order term $T_s^{(1)}$ from Eq.~(\ref{eq:T1_energy}) contains the operator $( -\hat{t} \hat{\rho}_B - \hat{\rho}_B \hat{t})$, which is very similar to the projector by Huzinaga and Cantu~\cite{huzinaga1971},
\begin{equation} \label{eq:huzinaga_cantu_operator}
	\hat{O}^{\mathrm{HC}} =  -\hat{f} \hat{\rho}_B - \hat{\rho}_B \hat{f},
\end{equation}
but features  the one-electron kinetic energy operator $\hat{t}$ instead of the Fock operator $\hat{f}$. One might find it surprising as the operator from Eq.~(\ref{eq:huzinaga_cantu_operator}) is often employed in projection-based embedding~\cite{manby2012,tamukong2014,chulhai2017} to enforce external orthogonality between subsystem orbitals, and therefore ensures that $T_s^{\mathrm{nad}} [\rho_A, \rho_B ] = 0$, whereas no external orthogonality requirements were adopted in our derivations. However, a close relation between the Huzinaga building-block equations for many-electron
systems~\cite{huzinaga1971} and the Adams--Gilbert formalism~\cite{adams1961,gilbert1964}, which similarly to the present work employs Slater determinants composed of non-orthogonal MOs, is known and was previously discussed in the literature~\cite{francisco1992}. Moreover, the expression from Eq.~(\ref{eq:T1_energy}) was derived in Ref.~\cite{francisco1992} for the more general case of one-electron operators.

To self-consistently employ the approximation from Eq.~(\ref{eq:final_non_additive_correction}) 
within sDFT, the corresponding potential $\upsilon^{\mathrm{nad}}_{\mathrm{kin}}$ has to be derived as well. As seen from Eq.~(\ref{eq:kin_func_der}), this requires computations of the functional derivative $\delta T_s^{\mathrm{nad}} [\rho_A] / \delta \rho_A  (\vec{r}\,)$ or, equivalently, derivatives of the expansion terms $T_s^{(n)}$. Since $T_s^{(n)}$ depend on MOs and are not known as explicit functionals of the density, these evaluations could be performed by using the Optimized Effective Potential method~\cite{sharp1953,talman1976,engel_dreizler2011}, which is, however, computationally very demanding. Instead, we follow the idea behind the Generalized Kohn--Sham (GKS) approach~\cite{seidl1996}, where the use of orbital-dependent energy contributions naturally results in orbital-dependent potentials. In other words, the action of the potential $\upsilon^{\mathrm{nad}}_{\mathrm{kin}}$ on an active subsystem MO $\psi^{A}_l$ is represented as the functional derivative of $T_s^{\mathrm{nad}}$ with respect to the complex conjugate $\psi^{A*}_l$~\cite{engel_dreizler2011}, i.e.,
\begin{equation} \label{eq:GKS_in_nutshell}
	\upsilon^{\mathrm{nad}}_{\mathrm{kin}} (\vec{r}\,) \psi^{A}_l (\vec{r}\,) \to [\upsilon^{\mathrm{nad}}_{\mathrm{kin}} \psi^{A}_l] (\vec{r}\,) = \frac{\delta T_s^{\mathrm{nad}} [\{ \psi^A_i \}, \{ \psi^B_i \}]}{ \delta \psi^{A*}_l}.
\end{equation}
For a more rigorous introduction of GKS, we refer to the original work in Ref.~\cite{seidl1996}. Note that expressions of the GKS theory were also formulated within the FDE formalism~\cite{laricchia2010}.  
Therefore, derivations of functional derivatives of the form $\delta T_s^{(n)} [\{ \psi^A_i \}, \{ \psi^B_i \}] / \delta \psi^{A*}_l$ for the first few expansion terms of Eq.~(\ref{eq:final_non_additive_correction}) are required.
These detailed derivations are presented in Sec.~S2 in the SI, whereas working equations in atomic orbital representation and description of the associated computational cost are given in Sec.~S4.1. The final expressions in the MO representation read
\begin{equation} \label{eq:func_der_T1}
	\frac{\delta T_s^{(1)} [\{ \psi_i^A \}, \{ \psi_i^B \}]}{\delta \psi_l^{A*}} = -\left( \hat{\rho}_B \hat{t}  +  \hat{t} \hat{\rho}_B   \right) \psi_l^A
\end{equation}
and 
\begin{equation} \label{eq:func_der_T2}
	\frac{\delta T_s^{(2)} [\{ \psi_i^A \}, \{ \psi_i^B \}] }{\delta \psi_l^{A*}} = \left(   \hat{\rho}_B \hat{\rho}_A \hat{t}  + \hat{t} \hat{\rho}_A \hat{\rho}_B + \hat{\rho}_B \hat{t} \hat{\rho}_B   \right) \psi_l^A.
\end{equation}
As one can see, the same operator $(-\hat{\rho}_B \hat{t} - \hat{t} \hat{\rho}_B)$ appears in the expressions for the first-order energy correction from Eq.~(\ref{eq:T1_energy}) and for the corresponding functional derivative from Eq.~(\ref{eq:func_der_T1}). However, a slightly different operator expression is found for the second-order expansion terms when compared to the corresponding energy expressions. The operators from Eqs.~(\ref{eq:func_der_T1}) and (\ref{eq:func_der_T2}) are used in the following as parts of the Fock operator to self-consistently account for the non-additivity of kinetic energy, whereas Eqs.~(\ref{eq:T1_energy}) and (\ref{eq:T2_energy}) are employed for energy evaluations. The latter fact distinguishes our theory from projection-based embedding, where the corresponding energy contributions are equal to zero by definition due to the enforced external orthogonality of subsystem orbitals, i.e., $T_s^{\mathrm{nad}} [\rho_A, \rho_B ] = 0$. Note also that the terms from Eqs.~(\ref{eq:func_der_T1}) and (\ref{eq:func_der_T2})
can be implemented such that the computational cost is cubic with respect to the number of active system basis functions. For more details on this topic, we refer to Sec.~S4.1 in the SI.

Despite the fact that the approximate non-additive kinetic energy $T_s^{\mathrm{nad}} [\rho_A, \rho_B ]$ is the main error source in sDFT computations (for example, see Ref.~\cite{arti2015}), the overall performance of the method also depends on the chosen XC functional. In practical computations, the so-called conjoint functionals~\cite{leelee1991,march1991}, a pair of XC and kinetic energy functionals sharing 
the same form of enhancement factor, are often applied. In this regard and in the context of this work, it was unclear whether development of new approximations for the non-additive kinetic energy alone would lead to inconsistencies in evaluating the embedding potential. Therefore, similar approximations were derived for the non-additive XC contributions in Secs.~S3 and S4.2 in the SI and tested in Sec.~\ref{sec:beh2}.

\section{Computational Details} \label{sec:comput_details}

\begin{figure}[!h]
	\centering
	\includegraphics[width=0.92\textwidth]{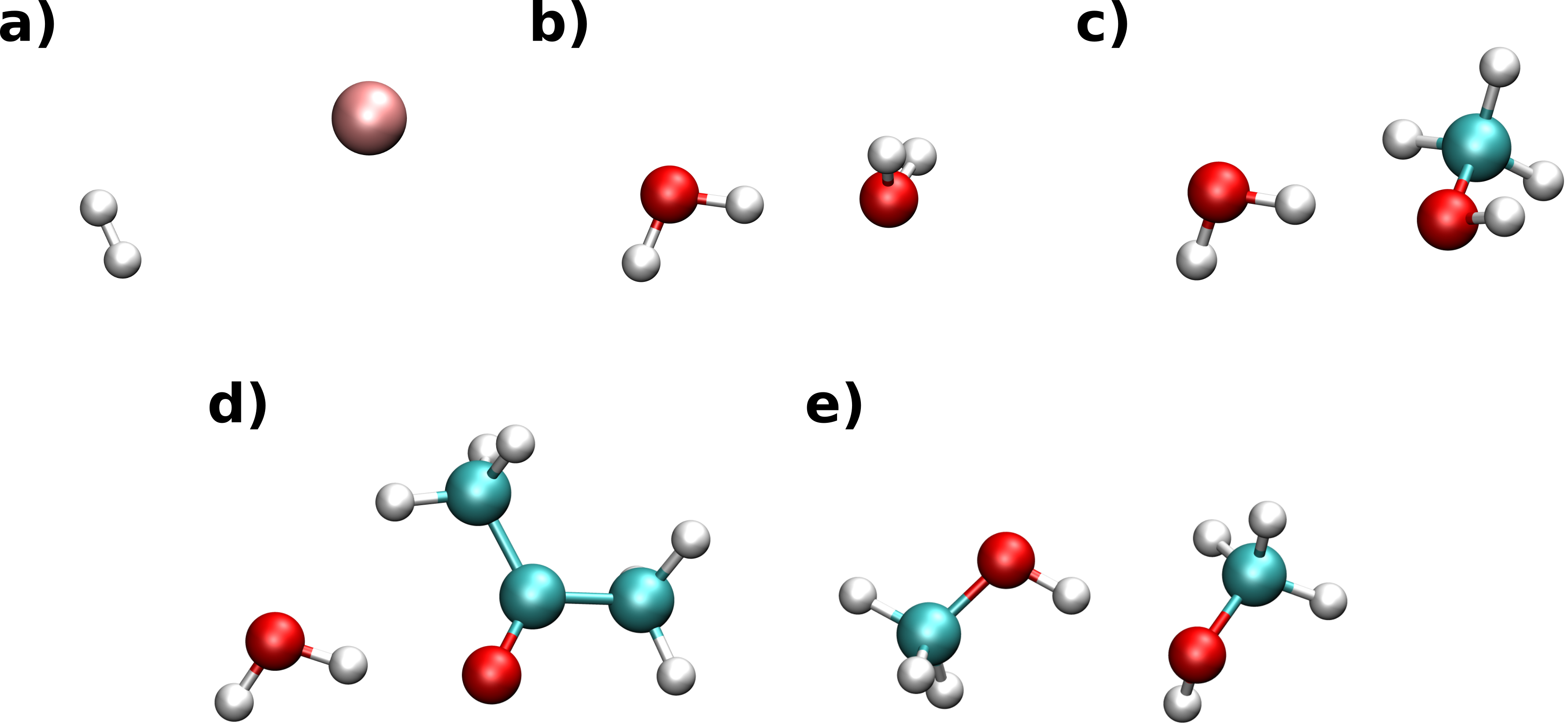}
	\caption{Molecular structures of a) Be$^+\cdots$H$_2$, b) H$_2$O$\cdots$H$_2$O, c) H$_2$O$\cdots$CH$_3$OH, d) H$_2$O$\cdots$(CH$_3$)$_2$O, and e) CH$_3$OH$\cdots$CH$_3$OH studied in this work.}
	\label{fig:structures}
\end{figure}

All computations presented in this work were carried out in a locally modified version of the \textsc{Serenity} program~\cite{unsleber2018,niemeyer2023,barton2024}.
The geometry of the T-shaped Be$^+\cdots$H$_2$ electrostatic complex was taken from Ref.~\cite{arti2014} and used without further structure optimization.
Molecular clusters of small solvent molecules such as water$\cdots$water (H$_2$O$\cdots$H$_2$O), water$\cdots$methanol (H$_2$O$\cdots$CH$_3$OH), water$\cdots$acetone  (H$_2$O$\cdots$(CH$_3$)$_2$O), and \linebreak methanol$\cdots$methanol  (CH$_3$OH$\cdots$CH$_3$OH) were optimized with KS-DFT using the PW91 XC functional~\cite{perdew1991,perdew1992} and the valence triple-$\zeta$ polarization def2-TZVP basis set~\cite{weigend2003,weigend2005}. The resulting molecular structures are shown in Fig.~\ref{fig:structures}. 
Subsequently, sets of displaced structures were created by varying the intermolecular O$\cdots$H and Be$^+\cdots$H$_2$ bond distances while keeping other degrees of freedom fixed. No further structure optimization was performed on resulting geometries. 

Generated molecular structures were used in subsequent KS-DFT and sDFT single point calculations.
To that end, the same def2-TZVP basis set~\cite{weigend2003,weigend2005} was employed for all molecular clusters except for Be$^+\cdots$H$_2$, which was computed using a smaller
3-21G basis~\cite{binkley1980}, unless stated otherwise. The XC PW91 functional~\cite{perdew1991,perdew1992} was consistently applied in both KS-DFT and sDFT computations. We defined and tested several different sDFT computational protocols varying in (i) the choice of the electron density ($\rho_{\Phi}$ or $\rho$) used for the evaluation of non-additive XC contributions, (ii) the non-additive kinetic energy approximation employed, and (iii) the truncation level $M$ of the Neumann expansion. Thus, standard sDFT computations using PW91~\cite{perdew1991,perdew1992} and PW91k~\cite{lembarki1994} functionals to account for non-additive contributions are referred to as sDFT/PW91k in the following. sDFT/EA[$M$] denotes computational protocols, which employ orbital-dependent embedding approximations of up to $M$th order of the Neumann series for the non-additive kinetic energy. If additionally the electron density $\rho_{\Phi}$ obtained from the $M$th order truncation of the series from Eq.~(31) from the SI is applied to evaluate non-additive XC contributions in conjunction with the PW91 XC functional, the notation sDFT/EA[$M$,$M$] is used. In all these cases, fully self-consistent computations were carried out and freeze-and-thaw cycles~\cite{weso1996} were performed until full convergence of electron densities, i.e., until the sum of absolute element-wise differences between density matrices from subsequent cycles was below the default convergence threshold of 1.0e$-$6~a.u.

For KS-DFT potential energy curves presented in Secs.~\ref{sec:beh2} and \ref{sec:add_examples}, the counterpoise (CP) correction scheme by Boys and Bernardi~\cite{boys1970} was used to account for the basis set superposition error. However, this error was not accounted for in sDFT-type computations and a monomer basis set was consistently applied in all cases. This is due to the fact that sDFT is reported to be free of the basis set superposition error unless charge-transfer-like interactions become important~\cite{dulak2007,Beyhan2013}. Furthermore, the goal of this work is to construct practical and computationally feasible approximations applicable to large molecular systems, which requires the use of a monomer basis set. Therefore, sDFT-type potential energy curves were computed according to the expression,
\begin{equation}
	\Delta E^{\mathrm{sDFT}} = E^{\mathrm{sDFT}} (AB; R) - E^{\mathrm{KS-DFT}} (A) - E^{\mathrm{KS-DFT}} (B),
\end{equation}
where $E^{\mathrm{sDFT}} (AB; R)$ is the sDFT energy of the complex $A\cdots B$ at the intermolecular distance $R$, while $E^{\mathrm{KS-DFT}} (A)$ and $E^{\mathrm{KS-DFT}} (B)$ are KS-DFT energies of isolated subsystems $A$ and $B$ in vacuum, respectively. Note that in the following, we refer to $\Delta E^{\mathrm{sDFT}}$ as to sDFT interaction energy.

To provide a quantitative measure for a difference between KS-DFT electron densities and those generated with other computational approaches, densities were first represented on 
accurate atom-centered Becke grids~\cite{becke1988,treutler1995,franchini2013} of the same size. For this purpose, integration grids of the highest quality from those available in the \textsc{Serenity} package were constructed (``accuracy 7''). By integrating grid-represented densities over the whole space and comparing results with the exact number of electrons, integration errors were found to be below about 2e$-$3 a.u. Then, absolute differences of a target density $\rho^{X} (\vec{r}\,)$ generated with sDFT-based protocols, from the reference KS-DFT results $\rho^{\mathrm{DFT}} (\vec{r}\,)$ on grids points $\vec{r}_i$ were computed and subsequently integrated over space, i.e.,
\begin{equation} \label{eq:int_dens_error}
   \sum_{i=1}^{N_{\mathrm{points}}} | \rho^{\mathrm{KS-DFT}} (\vec{r_i}\,)  -  \rho^{X} (\vec{r_i}\,)  | \, \omega_i,
\end{equation}
where $\omega_i$ are integration weights for grid points $\vec{r}_i$. A similar grid-based integration technique was used to verify whether the density $\rho_{\Phi}$ from Eq.~(31) from the SI integrates to the correct number of electrons.

\section{Results} \label{sec:results}

In what follows, we first analyze the convergence of the Neumann series for a number of molecular clusters in Sec.~\ref{sec:results_neumann}. To that end, overlap matrices generated with standard sDFT computations employing density-dependent approximations for the non-additive kinetic energy are used. Then, in Sec.~\ref{sec:beh2} the performance of newly proposed approximations is analyzed in detail for the test case of Be$^+ \cdots$H$_2$ electrostatic complex. Finally, in Sec.~\ref{sec:add_examples} a semi-empirical approach to calculating interaction energies is proposed and demonstrated.  

\subsection{Convergence of the Neumann Series} \label{sec:results_neumann}

A necessary and sufficient condition for the convergence of the Neumann series from Eq.~(\ref{eq:overlap_neumann}) is that the spectral radius $R$ of the matrix $ \mathbf{A} = (\mathbf{I}-\mathbf{S})$, i.e., the largest absolute eigenvalue $\lambda_i$ of $\mathbf{A}$,	
\begin{equation}\label{eq:spectral}
	R(\mathbf{A}) = \max_{i}\vert \lambda_i\vert,
\end{equation}
is smaller than one~\cite{renardy2004}. Unfortunately, a formal mathematical proof of convergence for general matrices of the form $(\mathbf{I}-\mathbf{S})$ cannot be given, as can be seen in the following example. Let us consider a helium dimer He$\cdots$He composed of two subsystems, which are labeled $A$ and $B$ and contain one helium atom each. In the case of restricted sDFT, the subsystem density $\rho_I (\vec{r}\,)$, where $I=A$ or $B$, is defined by the corresponding doubly occupied MO $\psi^I$. The inter-subsystem overlap integral is then equal to $s:=\langle \psi^A|\psi^B\rangle=\langle\psi^B|\psi^A\rangle$ and the $[2\times2]$ matrix $\mathbf{A} = (\mathbf{I} - \mathbf{S})$ is given by
 \begin{equation}\mathbf A =
 	\begin{pmatrix}
 		0 & -s \\
 		-s & 0
 	\end{pmatrix}.
 \end{equation}
The eigenvalues of this matrix can be found analytically and are equal to $\pm s$. Therefore, the spectral radius $R(\mathbf{A})$ of $\mathbf{A}$ is equal to the absolute value $|s|$ of $s$ and is smaller than or equal to one. This means that for the molecular system considered the Neumann series converges for all values $|s|\in[0,1)$ and diverges for $|s|=1$. The divergent case, however, corresponds to the nuclear fusion of two helium atoms and is of no concern for any realistic chemical system.

\begin{figure}[!h]
	\centering
	\includegraphics[width=0.75\textwidth]{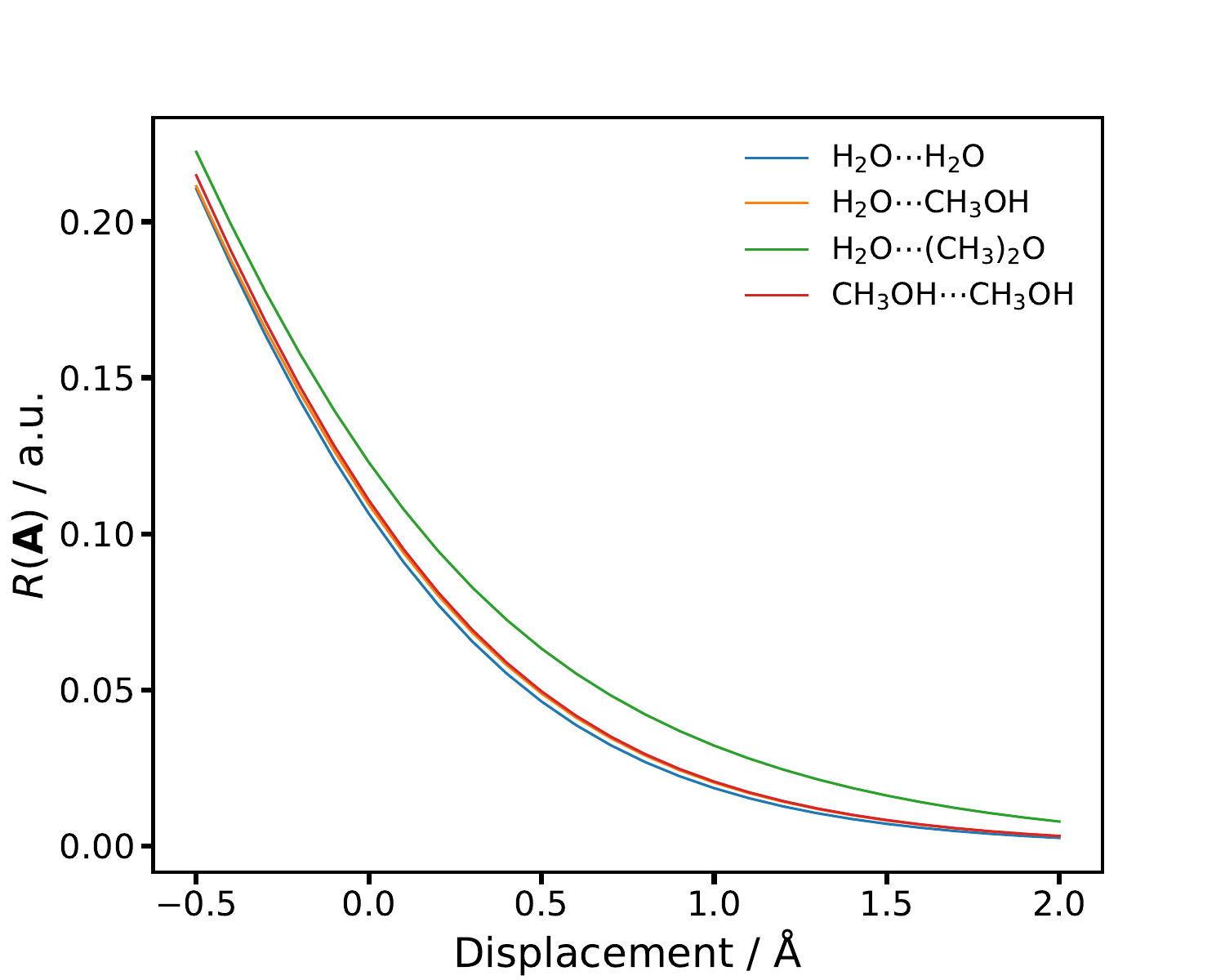}
	\caption{Spectral radii $R(\mathbf{A})$ for complexes H$_2$O$\cdots$H$_2$O, H$_2$O$\cdots$CH$_3$OH, H$_2$O$\cdots$(CH$_3$)$_2$O, and CH$_3$OH$\cdots$CH$_3$OH at different intermolecular displacements. Computations of overlap matrices were performed with sDFT/PW91k.}
	\label{fig:spectral-radii}
\end{figure}

For larger molecular systems, spectral radii $R(\mathbf{A})$ can be computed numerically. Such sDFT/PW91k computations for the molecular clusters H$_2$O$\cdots$H$_2$O, \linebreak H$_2$O$\cdots$CH$_3$OH, CH$_3$OH$\cdots$CH$_3$OH, and  H$_2$O$\cdots$(CH$_3$)$_2$O at different intermolecular displacements are presented in Fig.~\ref{fig:spectral-radii}. As one can see, the spectral radii $R(\mathbf{A})$ are well below one for all molecular clusters at all investigated displacements, therefore, signifying convergence of the Neumann series. Furthermore, it can be seen that the spectral radii $R(\mathbf{A})$ tend to zero for larger displacements. This is due to inter-subsystem overlap integrals tending to zero and, hence, $\mathbf A$ becoming the zero-matrix $\mathbf{0}$. 

As an alternative to direct and rather expensive numerical computations of eigenvalues, the Ger\v{s}chgorin circle theorem~\cite{gerschgorin1931} can be employed to evaluate an upper bound of spectral radii. In the general case, this theorem provides access to a set of disks in the complex plane, which contains eigenvalues of a matrix. Since we are dealing with symmetric, real-valued matrices $\mathbf A$, which have zeros as diagonal elements, the eigenvalues are contained within real-valued intervals centered at the origin $(0)\in\mathbb R$. The lengths of these intervals $l_i$ are equal to the sum of absolute values of elements belonging to $i$th row or column, i.e., for row-sums
\begin{equation}
	l_i = \sum_{k=1} | (\mathbf{A})_{ik} |.
\end{equation}
Hence, the convergence of the Neumann series is guaranteed in cases of Ger\v{s}chgorin intervals being in or equivalent to the interval $(-1,1)$. With inter-subsystem overlaps $s$ in $[-1,1]$, this holds true if all column or all row sums of the absolute values of entries of $\mathbf A$ are smaller than $1$. This sufficient condition offers a simple way to predict the convergence of the Neumann series in chemically relevant systems. 

In addition to the formal convergence of the Neumann series in the limit of an infinite number of expansion terms, its convergence rate is also of particular interest. Thus, if a considerably large number of expansion terms is required to approximate the inverse MO overlap matrix $\mathbf S^{-1}$, evaluations of $T_s^{\mathrm{nad}}$, as given in Eq.~(\ref{eq:final_non_additive_correction}), would become computationally very inefficient. Therefore, it is important to assess the performance of the Neumann series at different truncation levels $M$ and identify the minimal number of terms needed for reaching a specific accuracy. To that end, we re-write Eq.~(\ref{eq:overlap_neumann}) by taking the difference between the inverse MO overlap matrix $\mathbf S^{-1}$ and a truncation of its expansion and subsequently computing a matrix norm of the whole expression as
\begin{equation} \label{eq:series_norm}
	\Delta = \Vert \mathbf S^{-1}- \sum_{n=0}^{M} (\mathbf{I} -\mathbf{S})^n \Vert.
\end{equation}
Here, $\Delta$ is a scalar value representing the error of truncation at order $M$. For practical applications of Eq.~(\ref{eq:series_norm}), the exact inverse MO overlap matrix $\mathbf S^{-1}$ has to be available, which is not the case. We avoid this issue by calculating an approximate value of $\Delta$ using the Moore--Penrose pseudo inverse~\cite{moore1920,bjerhammar1951,penrose1955} of $\mathbf{S}$. Furthermore, different matrix norms can be applied in Eq.~(\ref{eq:series_norm}). In this work, we tested the performance of the 1-norm $\Delta_1$, 2-norm $\Delta_2$, $\infty$-norm $\Delta_{\infty}$, and Frobenius norm $\Delta_\mathrm{F}$~\cite{Higham2008}. Very similar results were obtained in all cases. Therefore, we limit our consideration here to only 2-norm $\Delta_2$. For more information on matrix norms, their properties, definitions as well as additional numerical tests, see Sec.~S5 in the SI.

\begin{figure}[!h]
	\centering
	\includegraphics[width=0.95\textwidth]{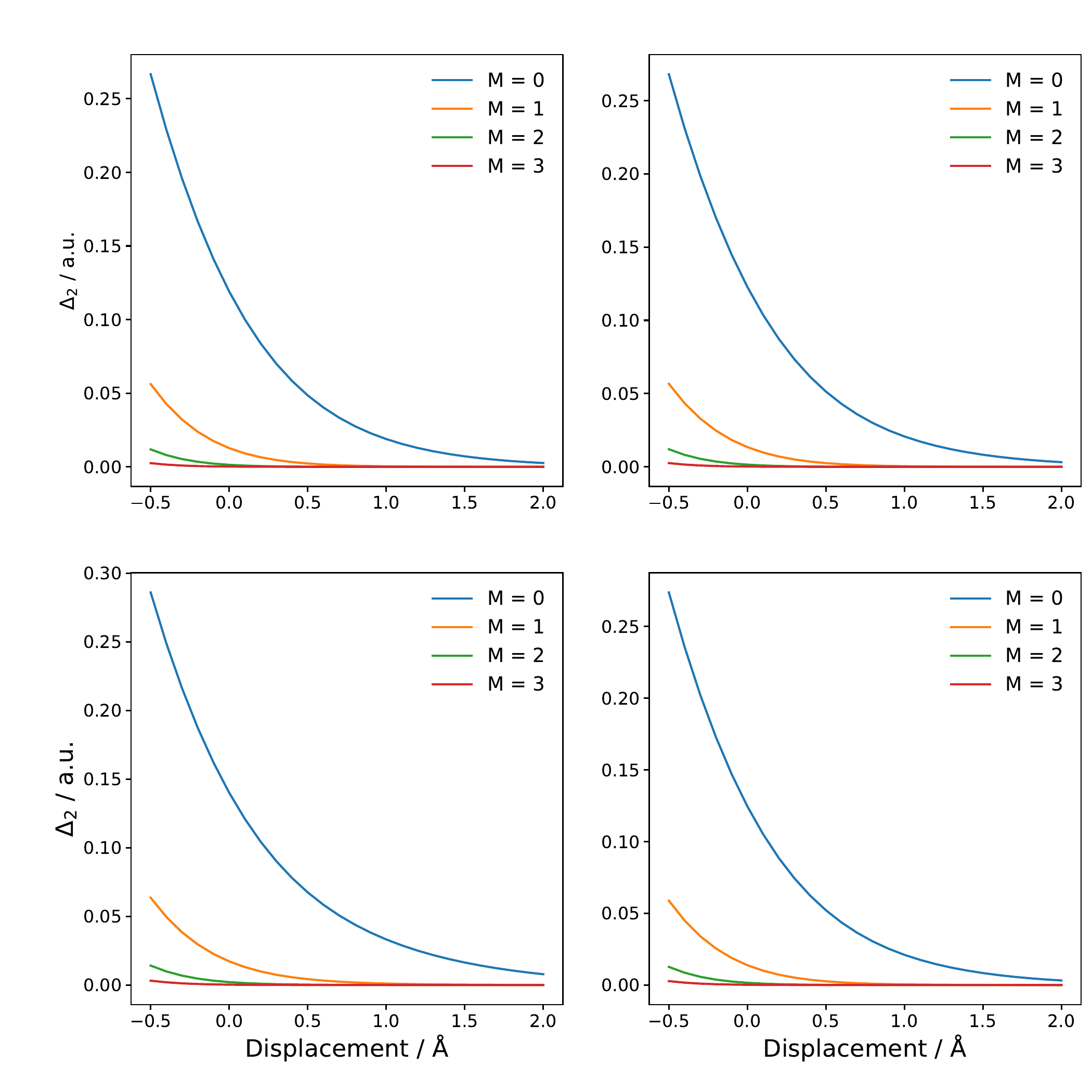}
	\caption{The error $\Delta_2$ computed at truncation levels $M=0,1,2$, and 3 of the Neumann series. The results are shown for molecular clusters of H$_2$O$\cdots$H$_2$O (top, left), H$_2$O$\cdots$CH$_3$OH (top, right), H$_2$O$\cdots$(CH$_3$)$_2$O (bottom, left) and  CH$_3$OH$\cdots$CH$_3$OH (bottom, right)
	at different intermolecular displacements relative to the equilibrium structure. Computations of overlap matrices were performed with sDFT/PW91k.}
	\label{fig:2-norms}
\end{figure}

Calculations of the truncation error $\Delta_2$ from Eq.~(\ref{eq:series_norm}) for a set of molecular complexes at different intermolecular displacements relative to the equilibrium structure are demonstrated in Fig.~\ref{fig:2-norms}. As can be seen, very similar results are obtained for all complexes. At intermolecular separations larger than about 2~\AA, the overlap matrix $\mathbf{S}$ and its inverse $\mathbf{S}^{-1}$ become identity matrices $\mathbf{I}$. This situation is accurately described by the Neumann series truncated at the zero order $M=0$, since the corresponding expansion term $(\mathbf{I} - \mathbf{S})^0$ is also equal to $\mathbf{I}$, whereas all higher-order terms $M>0$ yield zero matrices $\mathbf{0}$. As the result, the error $\Delta_2$ is equal to zero. At shorter displacements, orbital overlaps grow and $\mathbf{S}^{-1}$ start deviating from $\mathbf{I}$. Therefore, the zero-order expansion term is no longer sufficient for describing $\mathbf{S}^{-1}$. Using the Neumann series truncated at the first order $M=1$, the error $\Delta_2$ can be kept around zero for intermolecular displacements as short as 0.5~\AA. However, higher expansion terms are required for even shorter distances. We find the second expansion order $M=2$ sufficient for our applications as it yields very small errors at the equilibrium distance and is less computationally demanding than the third-order expanded series.

\subsection{Non-Additive Corrections}\label{sec:beh2}

For the initial numerical testing of our new approach, we employ a small electrostatic complex of the beryllium cation Be$^+$ with a hydrogen molecule H$_2$. This complex was previously investigated theoretically (for examples, see Refs.~\cite{poshusta1971,hinze1999,page2010,arti2014,dawoud2023}) and experimentally~\cite{roth2006}. It is known that Be$^+$ and H$_2$ are bound by weak ($D_e \approx$  0.4 eV) electrostatic and induction interactions resulting in a T-shaped molecular geometry. The ground and first excited electronic states are well-separated from each other allowing us to apply single-reference electronic-structure methods. These make it a convenient example to test the performance of sDFT-based approaches. In fact, this compound was previously employed for numerical tests in Ref.~\cite{arti2018}.

Before presenting results generated with the new computational scheme, we point out two potential issues. First, by introducing new correction terms dependent on the non-orthogonal MOs, we, in principle, incorporate a new electron density $\rho_{\Phi} (\vec{r}\,)$ in sDFT as seen from Eq.~(31) from the SI. This new density is not equal to $\rho (\vec{r}\,)$, which is given in Eq.~(\ref{eq:dens_partitioning}) and is being optimized/relaxed within sDFT to find the minimum of energy. This creates an inconsistency in the overall approach and means that the new scheme can no longer be considered formally exact as opposed to standard KS-DFT and sDFT. The second issue stems from the fact that $\rho_{\Phi} (\vec{r}\,)$ does not necessarily integrate to the number of electrons $N$ in the total molecular system, when being represented as a truncated Neumann series. This is easy to see from Eq.~(31) in the SI, where the zero-order term $\rho^{(0)} (\vec{r}\,)$ is equal to the electron density $\rho (\vec{r}\,) = \rho_B (\vec{r}\,) + \rho_A (\vec{r}\,)$ and, therefore, by definition integrates to the number of electrons $N$. As a consequence, integration of $\rho_{\Phi} (\vec{r}\,)$ results in $N$ only if the integral of the sum of higher-order expansion terms is equal to zero. This requirement is satisfied for the infinitely large series, but does not necessarily hold for all possible truncated expansions.
To further analyze this aspect, we computed integrals of $\rho_{\Phi} (\vec{r}\,)$ with sDFT/EA[$M$] for different truncation orders $M$ as seen in Fig.~S2 in the SI. Our results demonstrate that the first-order truncated density expansion, $M=1$, does not correctly reproduce the number of electrons in Be$^+\cdots$H$_2$ for intermolecular displacements shorter than about 1.5~\AA. The deviation reaches about minus one electron for the equilibrium distance (i.e., at the displacement of 0.0 \AA). This also shows that the density correction $\rho^{(1)}$ could be negative. However, already for $M=2$ the correct number of electrons $N = 5$ is obtained for all displacements. Furthermore, no negative density areas were found when analyzing the second-order expanded $\rho_{\Phi}$. Higher-order terms were found to have negligible contributions to the number of electrons.

\begin{figure}[!ht]
	\centering
	\includegraphics[width=0.95\textwidth]{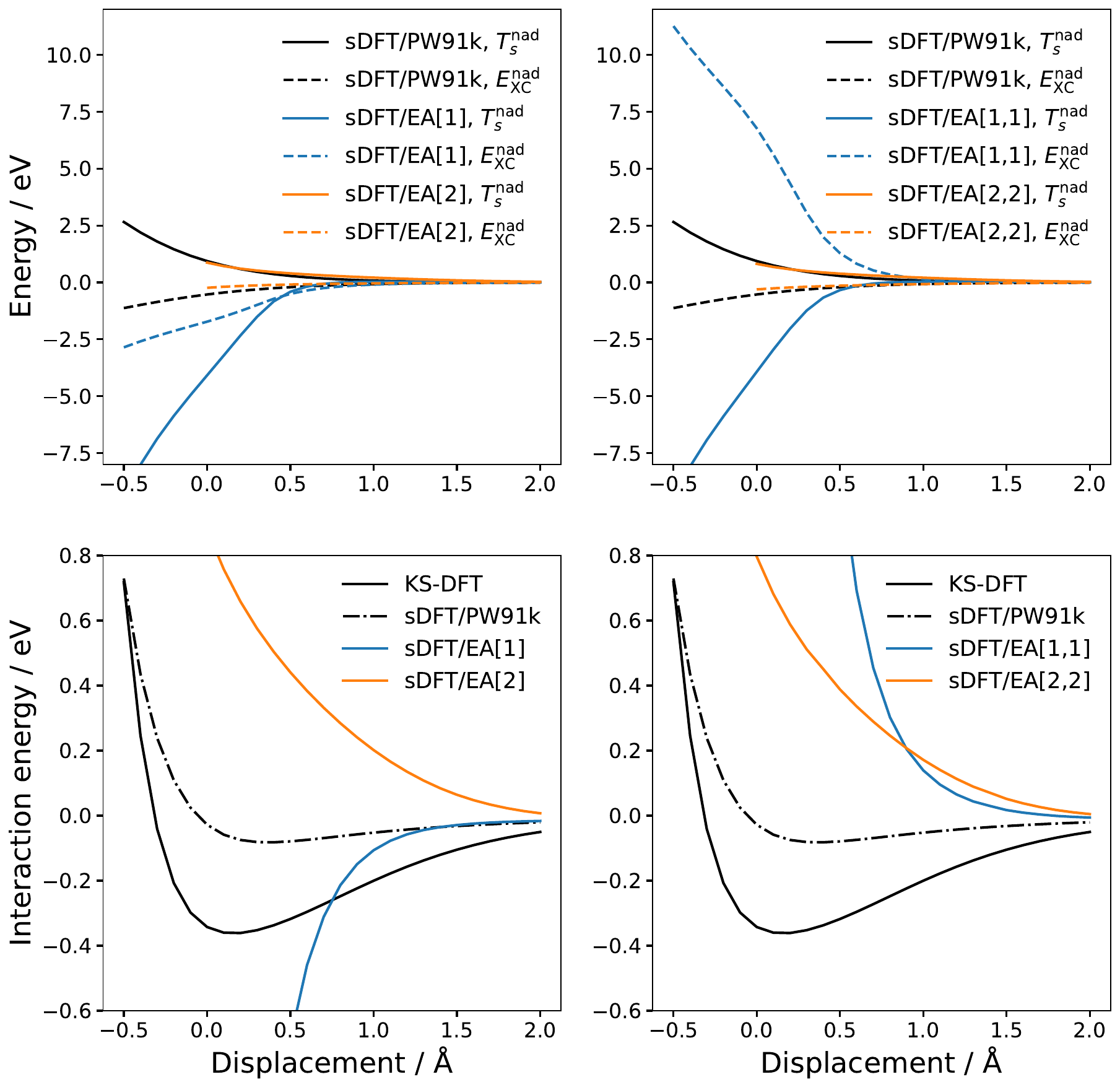}
	\caption{Non-additive energy contributions and interaction energies as functions of the intermolecular displacement computed for the Be$^+\cdots$H$_2$ complex with KS-DFT and sDFT-based approaches. Non-additive contributions computed with sDFT/EA[$M$] and sDFT/EA[$M$,$M$] are shown in top left and top right, respectively. Interactions energies obtained with sDFT/EA[$M$] and sDFT/EA[$M$,$M$] are given in bottom left and bottom right graphs, respectively. KS-DFT and sDFT/PW91k results serve as reference.}
	\label{fig:kin_nadd_en_corr}
\end{figure}

As the next step, we analyze the performance of new approximations by computing non-additive energy contributions and potential energy curves. To that end, KS-DFT and sDFT/PW91k approaches are used as reference. The results are shown in Fig.~\ref{fig:kin_nadd_en_corr}. As one can see from Fig.~\ref{fig:kin_nadd_en_corr} (top left), the non-additive energy contributions computed with sDFT/PW91k have opposite signs. The non-additive kinetic energy $T_s^{\mathrm{nad}}$ is always positive and, to a large extent, cancels negative $E_{\mathrm{XC}}^{\mathrm{nad}}$ contributions to the interaction energy. In fact, it was proven that the non-additive kinetic energy $T_s^{\mathrm{nad}}$ computed as a functional of electron density is always non-negative~\cite{weso2003}. The corresponding sDFT/PW91k potential energy curve is qualitatively correct, but strongly underestimates the interaction strength when compared to KS-DFT as seen from Fig.~\ref{fig:kin_nadd_en_corr} (bottom left).  Contrary to that, both sDFT/EA[1] non-additive energies are negative and result in qualitatively incorrect and quickly descending potential curves, see Figs.~\ref{fig:kin_nadd_en_corr} (top left) and (bottom left). This is probably a consequence of the first-order expansion term of the Neumann series from Eq.~(\ref{eq:T1_energy}) featuring a negative sign. This assumption is supported by sDFT/EA[1,1] computations, demonstrated in Figs.~\ref{fig:kin_nadd_en_corr} (top right) and (bottom right), which lead to both non-additive contributions having opposite signs to those from sDFT/PW91k. In this case, there is partial cancellation between non-additive contributions. However, the corresponding sDFT/EA[1,1] potential energy curve is overly repulsive and still qualitatively incorrect. 
Note that the sign of $T_s^{(1)}$ depends on the truncation level $M$ since our non-additive corrections are employed self-consistently. It is negative in sDFT/EA[1] computations, but becomes positive when higher-order correction terms are included, i.e., for sDFT/EA[$M$] with $M>1$. This fact is demonstrated in Sec.~S7 in the SI.
All sDFT/EA[2] and sDFT/EA[2,2] non-additive energy contributions show a much better agreement with sDFT/PW91k results, as seen from Figs.~\ref{fig:kin_nadd_en_corr} (top left) and (top right), respectively, and reproduce signs correctly. However, the deviations are still too large to correctly reproduce the shape of the associated potential energy curves. Additionally, sDFT/EA[2] and sDFT/EA[2,2] have convergence issues in self-consistent field procedures at intermolecular displacements shorter than 0.0~\AA. As can be seen from Figs.~\ref{fig:kin_nadd_en_corr} (bottom left) and (bottom right), both curves have a qualitatively incorrect non-bonding character. It should also be noted that potential energy curves and non-additive XC energy contributions $E_{\mathrm{XC}}^{\mathrm{nad}}$ are very similar for the sDFT/EA[2] and sDFT/EA[2,2] approaches. Therefore, the use of the electron density $\rho_{\Phi}$ for evaluations of the non-additive XC contribution does not significantly change the outcome of computations when second- or higher-order expansion terms are employed. Additionally, we assessed the performance of the sDFT/EA[3] computational protocol. The obtained interaction energies were found to be very similar to those from sDFT/EA[2] with the largest deviation of about 0.003~eV. Therefore, we conclude that the Neumann series is sufficiently well converged at the second order of truncation and the use of even higher-order terms is not likely to lead to considerable improvements. Finally, we analyzed basis set and XC functional dependencies of the sDFT/EA[$M$] and sDFT/EA[$M$,$M$] computational schemes. To that end, similar computations of interaction energies for Be$^+\cdots$H$_2$ were performed using the double-$\zeta$ def2-SVP and triple-$\zeta$ def2-TZVP~\cite{weigend2003,weigend2005} basis sets and 
pairs of XC and kinetic energy functionals such as (i) LDA  
and TF~\cite{Thomas1927,Fermi1928} and (ii) BP86~\cite{beck1988,perd1986} 
and LLP91K~\cite{leelee1991}. Results of this analysis are shown in 
Sec.~S8 in the SI. A strong dependency of standard sDFT computations on XC and kinetic energy functionals was reported before in the literature (for example, see Ref.~\cite{schluens2015}) and was, therefore, expected to be observed for sDFT/EA[$M$] and sDFT/EA[$M$,$M$] as well. However, qualitatively same and unsatisfactory results, to those presented in Fig.~\ref{fig:kin_nadd_en_corr}, were obtained with sDFT/EA[$M$] and sDFT/EA[$M$,$M$] showing only rather minor dependencies on the basis set and XC functional applied.

\subsection{Semi-Empirical Approach} \label{sec:add_examples}

\begin{figure}[!ht]
	\centering
	\includegraphics[width=0.99\textwidth]{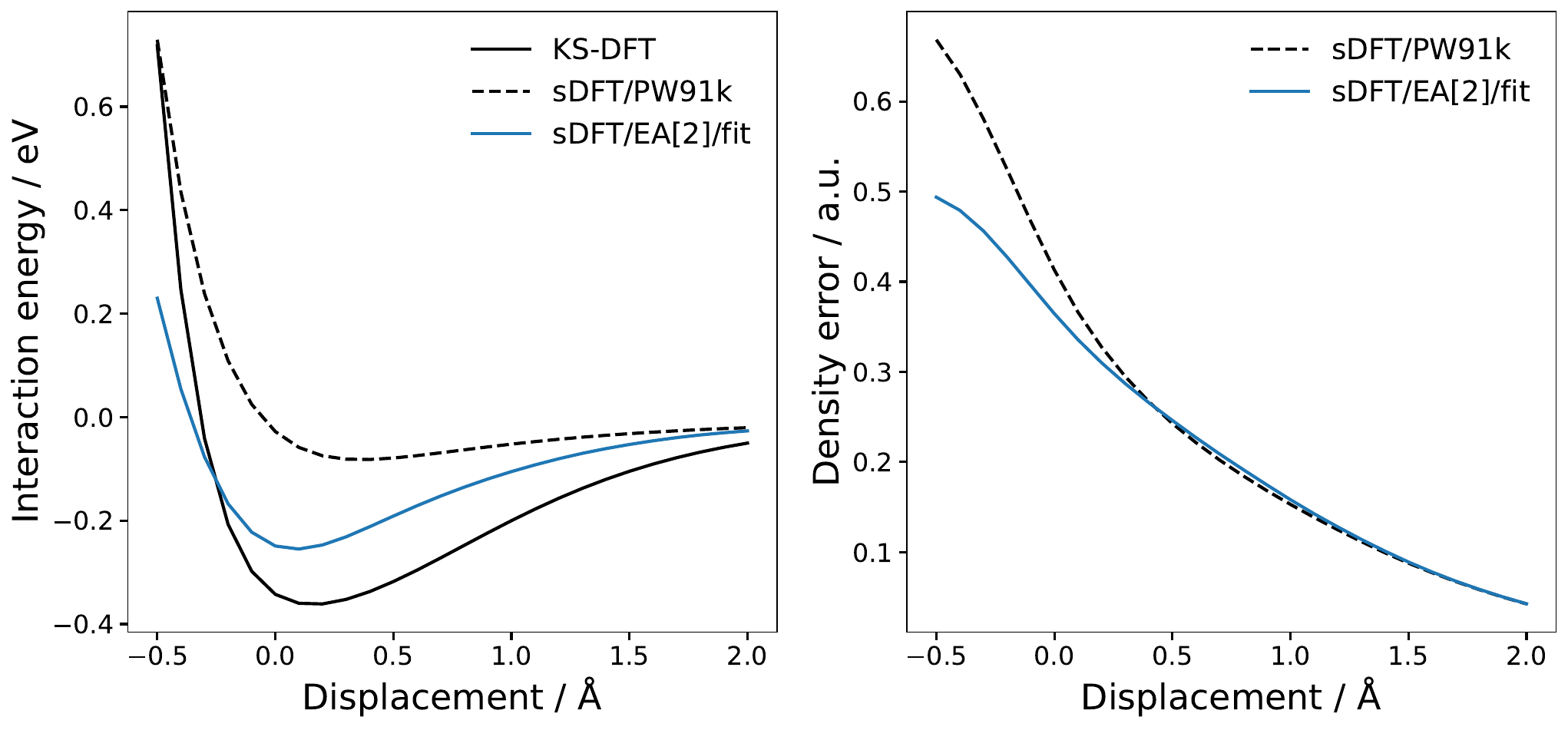}
	\caption{Interaction energies (left) and integrated density errors (right)  as functions of the intermolecular displacement computed for the Be$^+\cdots$H$_2$ complex with KS-DFT and sDFT-based approaches. sDFT density errors are computed according to Eq.~(\ref{eq:int_dens_error}) with KS-DFT density serving as the reference.}
	\label{fig:beh2_fitted}
\end{figure}

As demonstrated in Sec.~\ref{sec:beh2}, the use of our approximations constructed for the non-additive kinetic energy led to qualitatively incorrect results. Since it was also shown that the Neumann series converges sufficiently well already at the second order, errors in the interaction energy are probably due to other assumptions made. First, as outlined above in Sec.~\ref{sec:nadd_kin}, we assumed that the non-interactive kinetic energy of the total molecular system can be computed according to Eq.~(\ref{eq:main_assumption}). Second, and as was pointed out previously, we incorporated an inconsistency by introducing a new density $\rho_{\Phi} (\vec{r}\,)$. Analyzing these assumptions in more detail is not a trivial task, which clearly goes beyond the scope of the current work. Instead, we adopt a more pragmatic approach and show how qualitatively correct results could still be obtained by introducing purely empirical parameters in orbital-dependent expressions for the non-additive kinetic energy. This decision is motivated by sDFT/EA[$M$] results showing very little dependence on the basis set and XC functional used. Therefore, a set of parameters found for one molecular system, and a specific basis set and functional might be transferable to other cases. To analyze this hypothesis, several parametric forms based on Eq.~(\ref{eq:final_non_additive_correction}) were tested. Among those are the non-additive kinetic energies $T^{\mathrm{nad}}_s$ being represented as $\alpha ( T^{(1)} +  T^{(2)})$, $\alpha T^{(1)} + \alpha^2 T^{(2)}$, and $\alpha T^{(1)} + \beta T^{(2)}$. Note that computations were still performed self-consistently and parameters $\alpha$ and $\beta$ were used as scaling factors for the associated energy- and Fock-matrix contributions. Best results were obtained with the latter fit expression, setting $\alpha$ to $-1.0$ and finding $\beta$ by minimizing deviations between sDFT/EA[2] and KS-DFT interaction energy curves of Be$^+\cdots$H$_2$. To that end, the PW91 XC functional and 3-21G basis set were employed. Results of this minimization procedure (with $\beta = 0.17$) are shown in Fig.~\ref{fig:beh2_fitted}. It can be argued that the use of a positive $\alpha$ parameter is a more natural choice. However, since $T_s^{(1)}$ and $T_s^{(2)}$ obtained from sDFT/EA[2] are positive, as was mentioned before in Sec.~\ref{sec:beh2}, such a fit does not result in a bound state.

As can be seen from Fig.~\ref{fig:beh2_fitted} (left), the fitted sDFT/EA[2] interaction energy curve (denoted as sDFT/EA[2]/fit) shows qualitatively correct behavior and outperforms sDFT/PW91k in reproducing the well-depth $D_e$ value. The corresponding equilibrium distance is shorter than that from KS-DFT, but agrees well with that from the original coupled cluster computation from Ref.~\cite{arti2014}. This improvement in performance of sDFT/EA[2]/fit is, of course, not surprising since the fitting was performed on the very same molecular cluster. However, deviations from KS-DFT results are still considerable. At the displacement of $0.0$~\AA, the difference between sDFT/EA[2]/fit and KS-DFT is about 0.1~eV. It is also interesting to note that the integrated sDFT/EA[2]/fit density error, computed according to Eq.~(\ref{eq:int_dens_error}), is smaller than that from standard sDFT/PW91k at short intermolecular displacements as seen from Fig.~\ref{fig:beh2_fitted} (right). 

\begin{figure}[!ht]
	\centering
	\includegraphics[width=0.99\textwidth]{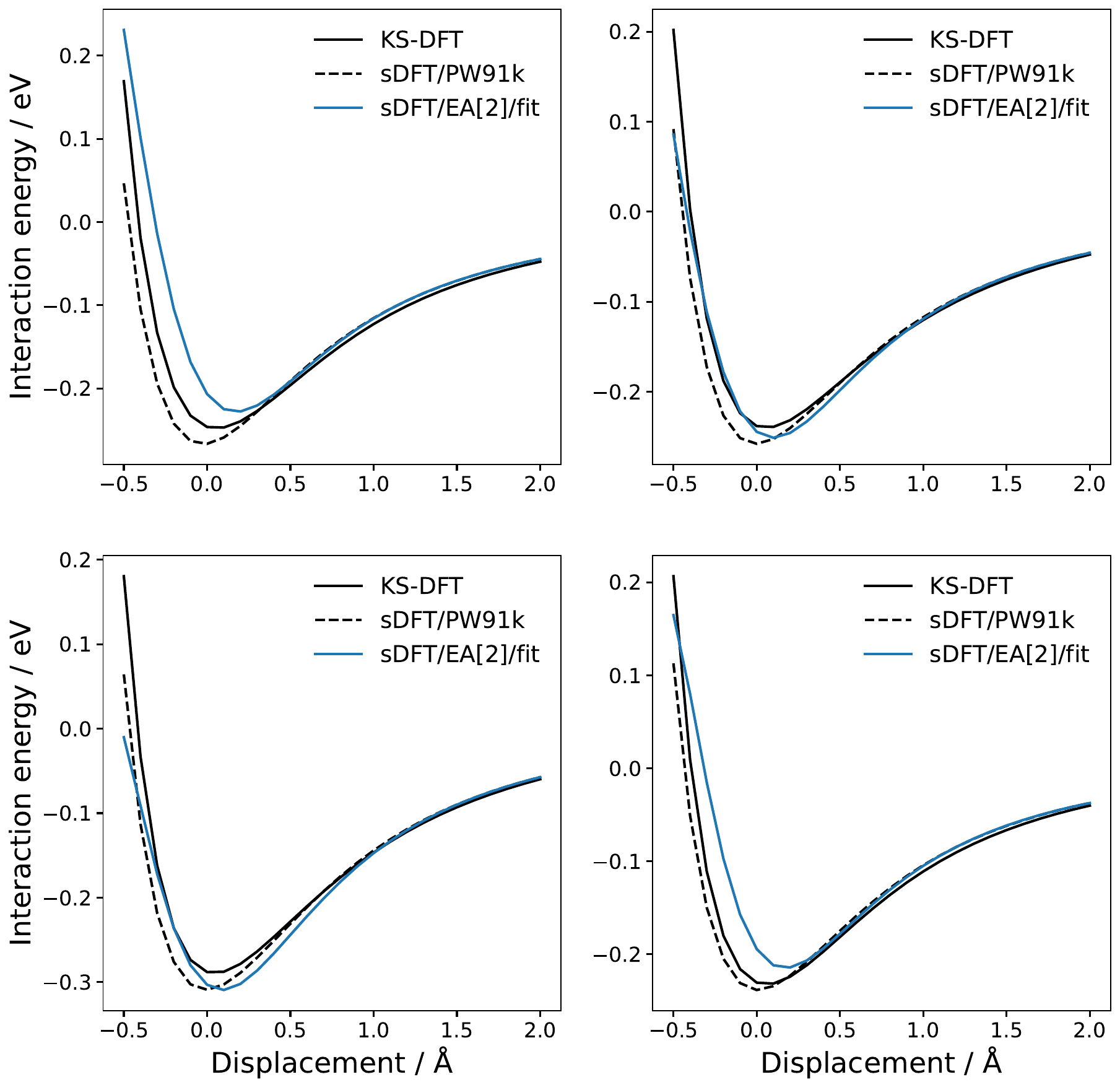}
	\caption{Interaction energies as functions of the intermolecular displacement computed for the H$_2$O$\cdots$H$_2$O (top left), H$_2$O$\cdots$CH$_3$OH (top right), H$_2$O$\cdots$(CH$_3$)$_2$O (bottom left), and CH$_3$OH$\cdots$CH$_3$OH (bottom right) molecular complexes with KS-DFT and sDFT-based approaches. KS-DFT and sDFT-based computations are performed using the PW91 XC functional and def2-TZVP basis set.}
	\label{fig:many_pes_fitted}
\end{figure}

Subsequently, the parameters found for Be$^+\cdots$H$_2$ were used without re-optimization for the H$_2$O$\cdots$H$_2$O, H$_2$O$\cdots$CH$_3$OH, H$_2$O$\cdots$(CH$_3$)$_2$O, and CH$_3$OH$\cdots$CH$_3$OH molecular complexes. To that end, the larger def2-TZVP basis set was employed. Results of these computations are shown in Fig.~\ref{fig:many_pes_fitted}. As one can see, all sDFT/EA[2]/fit interaction energy curves show qualitatively correct behavior. Furthermore, issues with converging the self-consistent field procedure are no longer observed. In all four cases, the use of sDFT/EA[2]/fit
results in slightly larger equidistant intermolecular distances than those from KS-DFT. Furthermore, the value of the well-depth $D_e$ is underestimated for H$_2$O$\cdots$H$_2$O and  CH$_3$OH$\cdots$CH$_3$OH by about 0.015~eV and overestimated for 
H$_2$O$\cdots$CH$_3$OH and H$_2$O$\cdots$(CH$_3$)$_2$O
by about 0.015~eV and 0.024~eV, respectively. Computing root-mean-square deviations of sDFT/PW91k and sDFT/EA[2]/fit interaction energies from reference KS-DFT results (on 26 equidistantly separated grid points for displacements from $-0.5$~{\AA}  to 2.0~\AA), we obtain errors below about 0.04~eV in all cases. sDFT/PW91k outperforms sDFT/EA[2]/fit for three compounds, namely 
H$_2$O$\cdots$H$_2$O, 
H$_2$O$\cdots$(CH$_3$)$_2$O, 
and CH$_3$OH$\cdots$CH$_3$OH, by only about 0.01~eV, whereas 
sDFT/EA[2]/fit shows a higher accuracy than sDFT/PW91k by 0.006~eV in the case of H$_2$O$\cdots$CH$_3$OH.
Furthermore, we computed sDFT/PW91k and sDFT/EA[2]/fit integrated density errors relative to KS-DFT. Results of this analysis are presented in Sec.~S9 in the SI. As one can see, sDFT/EA[2]/fit outperforms sDFT/PW91k in all cases except for the  H$_2$O$\cdots$H$_2$O complex.
We, therefore, conclude that the proposed semi-empirical approach is indeed robust and has potential to be transferable between different molecular systems and basis sets. However, a thorough benchmark study is required to find optimal parameters applicable to a broad range of molecular systems and interactions. In this regard, the extended molecular test sets S22x5~\cite{grafova2010} and S66x8~\cite{rezac2011} are especially attractive. Furthermore, other parametric forms could be investigated. This, however, goes beyond the scope of the current work and will be conducted elsewhere.

\section{Summary and Conclusions} \label{sec:conclusions}

In this work, we presented an alternative route to constructing inexpensive approximations for the non-additive kinetic energy contribution in sDFT. The use of Slater determinants composed of non-orthogonal Kohn--Sham-like MOs for computing the kinetic energy of the total molecular system as an expectation value and the Neumann expansion of the inverse MO overlap matrix constitute the core of this methodology. By deriving the first few terms of the Neumann-expanded kinetic energy expression and taking the corresponding functional derivatives, we constructed a series of orbital-dependent approximations to the non-additive kinetic energy, which can be directly and self-consistently incorporated in sDFT. We also pointed out and discussed differences and similarities of obtained expressions with those from the projection-based embedding theory employing the Huzinaga operator~\cite{huzinaga1971}. For testing purposes, similar approximations for the non-additive XC energy contributions were derived as well. It should be noted, however, that current derivations were carried out for the case of two subsystems and would require introducing additional approximations to be formulated in the general case of multiple subsystems.

Subsequently, we studied the behavior of the Neumann series in detail and discussed necessary and sufficient conditions for its convergence based on the eigenvalue analysis. For larger molecular systems, an alternative inexpensive technique for performing this analysis was proposed. Furthermore, we demonstrated that the Neumann expansion converges sufficiently well already at the second-order truncation level for molecular systems investigated in this work, namely water$\cdots$water, water$\cdots$methanol, water$\cdots$acetone, and methanol$\cdots$methanol clusters, and for a large range of intermolecular displacements. The inclusion of higher-order terms affected results slightly and was found to be important only for very short intermolecular displacements and strongly interacting molecular systems. Therefore, we conclude that the Neumann series is an efficient and robust tool for approximating the inverse MO overlap matrix and avoiding expensive matrix inversion operations.

The derived approximations were applied for computations of potential energy curves of the Be$^+\cdots$H$_2$ electrostatic complex and compared against standard KS-DFT and sDFT approaches, which employed explicit functionals of density.
Although corrections
to the non-additive kinetic and XC energies expanded to the second-order showed an agreement with the corresponding energy contributions from sDFT, the resulting potential energy curves were qualitatively incorrect. Inclusion of higher-order correction terms as well as the use of different XC functionals and basis sets did not lead to improved results. In fact, very little dependence of our computational approach on the choice of the XC functional and basis set was observed. 

This led us to the idea of introducing empirical parameters into the derived expressions and optimizing them such that deviations to KS-DFT potential energy curves for Be$^+\cdots$H$_2$ are minimized. As expected, the use of these new semi-empirical approximations resulted in improved accuracy and better agreement with the KS-DFT reference for the Be$^+\cdots$H$_2$ complex. Most importantly, however, we demonstrated that the very same parameters can be employed for calculations of other molecular clusters while using a larger basis set still resulting in quantitatively correct interaction energies. Thus, for water$\cdots$water, water$\cdots$methanol, water$\cdots$acetone, and methanol$\cdots$methanol complexes, the average deviations from KS-DFT energies were about 0.04~eV. For comparison, standard sDFT employing the decomposable PW91k~\cite{lembarki1994} approximation led to comparable accuracy and was even slightly outperformed by our semi-empirical approach in the case of the water$\cdots$methanol complex. Further, our approach demonstrated smaller integrated density errors than sDFT for all complexes except for water$\cdots$water.
Therefore, based on these proof-of-principle computations, we conclude that the obtained semi-empirical approximations have a potential to be transferable between different molecular systems. 

In conclusion, this work is an important step towards developing novel orbital-dependent approximations for the non-additive kinetic energy in sDFT. Although the current computational protocol requires the use of empirical parameters to correctly reproduce potential energy curves, it also shows a very good agreement with reference KS-DFT results for a set of molecular complexes, weak dependency on the basis set and XC functionals employed, and a high potential of optimized parameters to be applicable to other types of chemical compounds without re-optimization. Furthermore, the use of these semi-empirical approximations comes with a cubic computational cost with respect to the number of atomic orbitals in the active subsystem. 
Therefore, this approach is only slightly more expensive than 
sDFT with explicit kinetic energy functionals and could be applied to large molecular systems. However, finding a suitable set of empirical parameters applicable to a broad range of molecular systems and interaction types could be a challenging task and requires a more thorough benchmarking, which will be conducted elsewhere.

\section*{Acknowledgements}
Funding was provided by the German Research Foundation (Deutsche Forschungsgemeinschaft, DFG) --- project number 545861628. L.S.E. thanks the German Academic Scholarship Foundation (Studienstiftung
des deutschen Volkes) for their support.
Helpful discussions with Dr. Jan P. G\"otze are greatly
acknowledged. The authors thank the HPC Service of FUB-IT, Freie Universität Berlin, for computing time.


\end{document}